\documentclass[acmsmall]{acmart}

\AtBeginDocument{%
  \providecommand\BibTeX{{%
    \normalfont B\kern-0.5em{\scshape i\kern-0.25em b}\kern-0.8em\TeX}}}

\acmJournal{JACM}

\begin{document}

\title[Coding with Purpose]{Coding with Purpose: Learning AI in Rural California}


\author{Stephanie Tena-Meza}
\email{tenameza@stanford.edu}
\affiliation{%
  \institution{Stanford University}
}

\author{Miroslav Suzara}
\email{[msuzara@stanford.edu}
\affiliation{%
  \institution{Stanford University}
}

\author{AJ Alvero}
\email{ajalvero@stanford.edu}
\affiliation{%
  \institution{Stanford University}
}

\renewcommand{\shortauthors}{Tena-Meza, Suzara, Alvero}


\begin{abstract}
We use an autoethnographic case study of a Latinx high school student from an agricultural community in California to highlight how AI is learned outside classrooms and how her personal background influenced her social-justice oriented applications of AI technologies. Applying the concept of learning pathways from the learning sciences, we argue that redesigning AI education to be more inclusive with respect to socioeconomic status, ethnoracial identity, and gender is important in the development of computational projects that address social-injustice. We also learn about the role of institutions, power structures, and community as they relate to her journey of learning and applying AI. The future of AI, its potential to address issues of social injustice and limiting the negative consequences of its use, will depend on the participation and voice of students from the most vulnerable communities.
\end{abstract}


\maketitle

\section{Introduction}

As artificial intelligence (AI)\footnote{We define AI here as data intensive computing \cite{dixon2019racializing}}. increasingly mediates everyday life, who gets to learn AI? What are the implicit goals and values of the people and organizations that currently have access to AI, and how do they translate into technology designs that impact society? As issues of fairness, ethics, and justice become increasingly germane to computer science (CS) education, understanding who has access to AI, their goals and values, and how their applications of AI could change society is necessary to enhance CS education and guide how and where efforts towards educational equity are focused. These questions are active areas of research in various CS communities, but the analysis is often centered around systems rather than the perspectives and experiences of individual people. Young people have wealths of knowledge about their communities that could inform defining and understanding of socially just applications of AI. By highlighting student perspectives of learning in these systems and communities, we can better understand their inner workings to make the learning CS and AI more just and socially aware.

Socially just applications of AI are not uniform: aside from certain overlapping issues, the most pressing problems in rural communities are not the same in urban communities. For example, agricultural migrant workers are exposed to deadly levels of pesticides and notoriously difficult working conditions \cite{osorio1998california}. Presumably, laborers and landowners have different conceptions about the utility and good that could come from AI which are shaped by their lived experiences and ideologies. However, these groups of people tend to be separated not just by the structure of US agriculture but also by race, social class, language, politics, citizenship, and opportunity \cite{shaw2010beyond}; we go into further detail below. Given these disparate circumstances and backgrounds, determining what socially just applications of AI could and should ``look'' like might depend on who is being asked to design and apply it and what they perceive to be in need of computational intervention. Well resourced businesses and organizations are able to pay for technological expertise to implement AI for their own goals, but asking students from marginalized communities what they envision and how they are learning CS and AI is crucial to shaping socially just applications of AI.

Recent learning sciences literature conceptualizes ``learning pathways'' as inclusive of place, time, and power dynamics \cite{nasir2020learningpathways}, and can serve as a powerful framework to explore this question. This paper presents a case study with retrospective autoethnographic components that describes the learning pathway of a Latinx high school student from an agricultural community in California. To do this, we devised a “reflective cycle” method where the focal student of the case study (an author of this paper) answered questions devised by the other authors and participate in discussion sessions where she reflected on her learning pathway. We show how her uses of AI to study water pollution from agriculture were shaped by her pathway, and how they run contrary to the automation-centric AI goals of powerful agricultural corporations in her hometown. For her, learning AI was not a platonic academic endeavor but rather was driven by her dedication to address the environmental injustices her family and community encounter as agricultural laborers. The future of AI education should consider how she and other students would address social injustice with computation.

Our paper makes the following contributions:
\begin{itemize}
    \item Illustrates how an individual’s identity influenced her applications of AI
    \item Highlights the benefits of reflective cycles for CS/AI education and research
    \item Shows how ``social good'' is defined and operationalized 
\end{itemize}

\section{Literature Review}

A wealth of research has investigated ways to broaden and democratize participation within computer science in the USA and abroad with respect to gender and race \cite{margolis2002unlocking,margolis2017stuck}. Recently, scholars and public figures have made a call for ``Computer Science for All’’\footnote{See \url{https://csforall.org/}} to reduce inequities and make learning more accessible in and out of schools \cite{goode2018computer,obama_2016}. Various equity-oriented strategies have arisen, but the work to truly carry out this vision remains ongoing and complex, as educators, policymakers, and various levels of government seek to align around what a shared vision and plan for action should look like \cite{ryoo2019going,vogel2017visions}. 

Despite these efforts, there remains a need to hear from students from underrepresented, marginalized, lower-income, POC communities \cite{ryoo2020minoritized}. Doing so would develop new understandings about broadening participation in CS and brings focus toward culture, identity, and lived experiences. Work by Moreno Sandoval applies a lens of Mexican ancestry epistemologies, shaped by epistemic theories and methods grounded in familial and ancestral forms of knowledge production, towards students learning CS in a southern California high school \cite{moreno2013critical,cruz2001toward}. Moreno Sandoval found that when tasked with applying computation to social problems, students relied heavily on their family experiences and knowledge in framing, analyzing, and describing the problem which in turn enhanced their learning experience and made it more significant. Engaging the cultural backgrounds of students learning AI could provide unique opportunities to reshape computational applications, their construction or dismantlement, of broader social systems and processes.

Thus, we draw upon the literature of culturally relevant and culturally responsive computing pedagogy \cite{madkins2019culturally} and more broadly culturally sustaining pedagogy \cite{alim2017culturally,paris2012culturally}. These asset-based approaches appreciate the identities and cultural knowledge of students while also recognizing, critiquing, and addressing current inequities within the system \cite{madkins2019culturally}. Relatedly, ‘community cultural wealth’ \cite{yosso2005whose} outlines the value of often overlooked assets to students of color, while ‘funds of knowledge’ \cite{moll1992funds} centers on the importance of home culture and lived experience. Combined, these frameworks are important to consider when thinking about who the learners are, and serve a critical piece to designing equitable and high-quality computer science education experiences. For example, the Exploring Computer Science program adopted a strategy of ‘Building on Students’ Funds of Knowledge and Cultural Wealth’ through placing computer science education ‘in the context of students’ communities, lived realities, and interests’ \cite{margolis2012beyond}.

Curriculum planning and development for K-12 AI is a growing area of CS education research \cite{touretzky2019envisioning}. However, research has identified various obstacles to implementation. First, data literacy is an area for growth across many sectors of K-12 education \cite{finzer2013data,shreiner2018data}. Access to technology, often called ``the digital divide'' is also a persistent issue \cite{warschauer2004technology}. There is even a slight terminological divide, as some of the curriculum development around data intensive computing has been co-occurring under the banner of ``data science’’ as opposed to AI \cite{gould2016teaching,mike2020data}. Regardless, researchers are also starting to think about relevant ethical concerns in K-12 education and curriculum \cite{dryer2018middle,ali2019constructionism}. The ongoing work in curriculum development is important, but higher order ecological and sociocultural learning frameworks are also necessary given the scope of social change instigated by AI. And, as AI becomes more directly linked with economic precarity as brought on by automation, giving marginalized students access and support to learning the technology is a minimum threshold to cross.

Fairness, ethics, and bias are becoming more central to AI education, and likewise there has been increased attention on applications of AI for ``social good''. The social harms of AI, especially related to race, socioeconomic status, gender, and other factors is well documented \cite{o2016weapons,noble2018algorithms,benjamin2019race}. Programs and frameworks are being developed at the university level to train people to use these technologies for social good (eg. \cite{floridi2020design}), but the projects and problems students work on tend to be defined from the perspective of organizations, businesses, and governmental entities rather than community members. How do students from different social backgrounds identify and define the problems whose solution would become a social good? There are persistent inequalities with computation and gender, but there have been fewer intersectional analyses that consider gender along with social class and/or race and how they might shape applications of AI. For example, a study showed bias in facial recognition for women but especially for Black women \cite{buolamwini2018gender}. Our conceptual framework explores how social good is and could be defined for K-12 AI education and prompts educators to focus on engaging students to help identify problems that could addressed with AI.

In making this literature review, we found that little has been written specifically about the experiences of AI education from the perspective of students. Studies have been published about elementary students learning AI models \cite{kim2017development} and many studies have analyzed the capabilities of AI powered tools in education. Our study expands this growing literature by focusing on K-12 AI education and describing in-depth the experiences of learning AI from one student’s perspective.

\section{Conceptual Framework}

We apply theories from the learning sciences, specifically the learning pathways framework, that explicitly consider systems of power in relation to student learning. We extend that literature to analyze the growing area of fairness, bias, ethics, and ``social good'' in AI education and their relationship to K-12 CS and AI education. This is particularly important because most K-12 students currently do not have clear pathways to learning AI. Social, cultural, political, and economic histories cannot be disentangled from applications of AI and that we should learn from students from marginalized backgrounds on how AI could/should be applied. Rather than assuming there is a universally agreed upon social good, we seek to understand what social good looks like from the perspectives of those not traditionally asked. 

\subsection{Learning Pathways}
Broadening participation in CS is a complex and multifaceted challenge that requires educators, researchers, and policymakers to go ‘beyond access’ in how we collectively think about inequity \cite{margolis2012beyond}. Similarly, this will also require to think about ``learning'' that situates individuals with respect to social systems, across time, across sites and geographies, and to enable individuals' participation in designing learning experiences. In this paper we conceptualize learning as inclusive of broader interactions with social others and learning environments. Through this lens, we consider ‘learning pathways’ as building upon ecological models of learning \cite{bronfenbrenner1979ecology} and extend that literature to the understudied area of pre-college AI education.

We center our theoretical grounding around the learning pathways \cite{nasir2020learningpathways} framework to account for learning in settings, across timescales, and inclusive of power dynamics. Three characteristics of learning pathways were particularly useful in framing our study: learning pathways are related to identities, made up of cultural practices and routines, and consider privilege and marginalization related to broader social and political institutions such as schools. Learning pathways extends the learning ecology framework by showing how hegemonic structures shape, mold, influence student learning experiences, identity-development, and continued participation based on their social, cultural, and economic backgrounds. We use these frameworks to interrogate the hegemonic structures in relation to AI learning systems in Salinas, California and consider individual and cultural factors. 
	 
\subsection{Defining Social Good}
People have begun to demand that social changes from AI address social good. But how is social good defined and operationalized? And who organizes the agenda for AI for social good? Defining social good and change in computing is a normative exercise \cite{abebe2020roles}, but the lack of gender and ethnoracial diversity in AI limits marginalized voices, opinions, and needs when setting agendas. We therefore adapt critical race theories in/of education to conceptualize social good in AI education \cite{margolis2012beyond,ladson2016toward} and argue that educators should use sociocultural, contextual frameworks for defining social good rather than assuming a positivist, universal existence of social good. This could help students make clearer connections to how this technology is influencing their individual lives in both beneficial and harmful ways.

Similar issues in defining social good as it relates to science and technology have been observed in medical grant-writing, for example. Grant topics about Black community health received lower scores than microscopic level research topics \cite{hoppe2019topic}. Community health and microbiology both address social good, but the racial and power asymmetries in agenda setting, which in turn define priorities for social good, should also be considered in AI education. Our paper addresses this by centering a young woman of color and her experiences learning AI in Salinas as well as her social-justice oriented applications of AI. Her experience highlights the importance of diversity in agenda setting for societal applications of AI \cite{benjamin2019race}. 

In agricultural communities, ideas of social good are similarly invoked through imagery and slogans about ``feeding America’’ and often belie the historical injustices faced by agricultural laborers. For example, a 2007 study found that traces of pesticides were found on many surfaces inside the homes of farmworkers \cite{bradman2007pesticides}, but many farmworkers across the US and California specifically do not have access to health insurance and/or healthcare \cite{chaney2017covariates}. The manual tools used by ag-workers have also been a source of controversy over the decades. One of the major legal victories of agricultural labor activist and organizer César Chávez was helping outlaw the short handled hoe, nicknamed \textit{el brazo del diablo} (``the devil's arm'') as it required users to bend over to cut weeds and vegetables and frequently caused back and skeletal injuries, in California during the 1970s \cite{cozzens2015defeating}. Dangerous pesticides and tools like the short handled hoe are used in agriculture because they are perceived by landowners to increase efficiency, quality, and their control over the agricultural process \cite{murray1982abolition}. As agricultural companies incorporate more AI through agricultural technology, or ag-tech (a term coined in Salinas \cite{gerty_2020}), there is a push towards automation (e.g. \cite{lakshmi2020artificial,kakani2020critical,guizzo2019your}) which in theory reduce labor costs by displacing farmworkers. The potential for AI to improve worker safety and/or reduce agricultural pollution has not been as central to ag-tech.

\begin{figure}[htb]
    \centering
    \includegraphics[width=\linewidth]{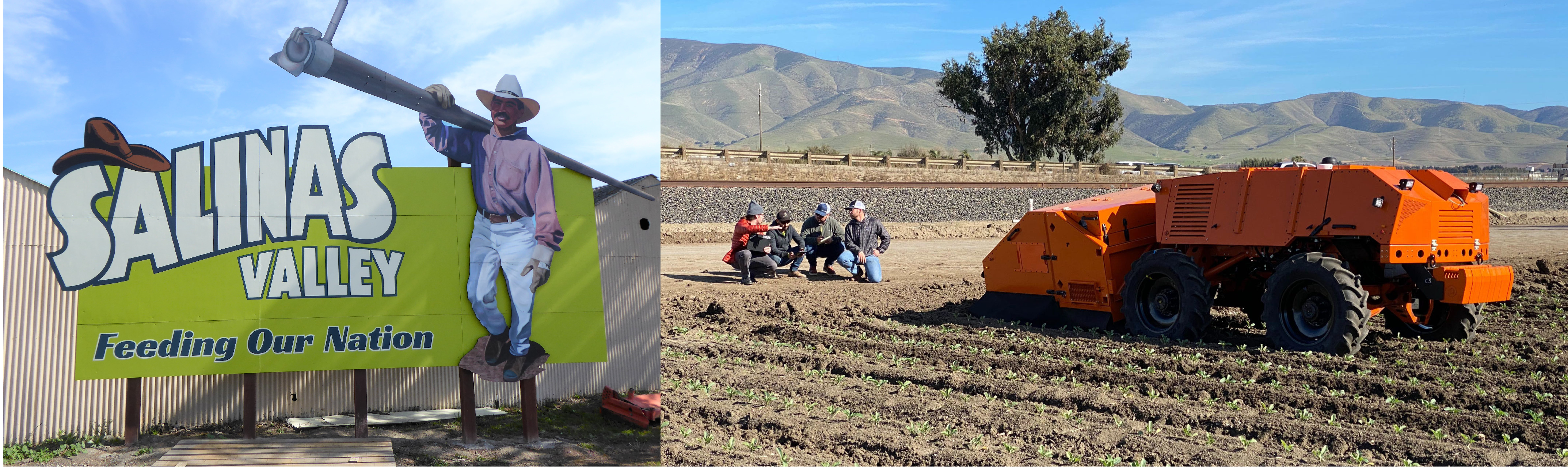}
    \caption{Envisioning the future of agricultural labor and technology in Salinas}
    \label{fig:future_salinas}
\end{figure}

The ``feeding America’’ slogans help shift focus away from the harms and negative ramifications of agricultural work and instead couch the rhetoric of agriculture in social good framings. Figure \ref{fig:future_salinas} highlights this by juxtaposing the social good in agriculture and ag-tech: On the left, a large mural reminds drivers entering Salinas from California Highway 68 that the city is ``feeding America’’ and the brown body doing the work. On the right, a Silicon Valley startup works with a small group of elites that run the multibillion dollar agricultural businesses in Salinas and uses AI and robotics to rapidly advance automation in agriculture to replace the actual people suggested by the figure on the left. Producing food for the country safely and efficiently is important, but much of the understanding of social good from the CS/AI community has been shaped through elite rather than community perspectives, and the harms experienced by farmworkers are not as central to social good conversations. CS education should therefore focus not just on inclusivity and access to this technology but also increase student awareness of the social ramifications of the tools they build and how they will address definitions of social good for some but not all. Without the changes to AI education we advocate, the future will look like the picture on the right: more large-scale automated ag-tech and more precarious and displaced farmworkers.

\subsection{Researcher Positionalities}
As researchers oriented toward a social justice framework who interrogate current educational ecosystems and access to CS and AI, we share and reflect upon our individual and collective identities \cite{muhammad2015reflections}. Two of the authors were born and raised in Salinas, the site of the study, and graduated from the same public high school. One identifies as a first-generation Mexican-American undergraduate contrasted with the other’s experiences as a first-generation Cuban-American doctoral student. The other author is a learning sciences PhD student who is not geographically connected with Salinas but is researching issues of equity, access, and social justice in CS.

\section{Methodology}

We present a case study with retrospective autoethnographic components of a first-generation Mexican-American high school student from Salinas, CA on her pathway of learning AI. This research builds upon the limited data that exists on K-12 AI education, especially from marginalized perspectives as they relate to race, social class, and/or gender. We chose to use a case study to learn about the student’s learning pathway because we wanted the student’s lived experiences to guide our research. The case study is based on the first author of this paper. The student’s recounting of stories of her first-hand experiences learning AI/CS outside the classroom was the primary method for generating the retrospective autoethnographic components in the case study. Her involvement throughout the research process has allowed our team to work collaboratively to develop this paper and gave the student an additional layer of agency in her participation.

To help illustrate the case study our team developed a ``reflective cycle'' which consisted of an iterative three-stage process: 1) brainstorming open-ended questions, 2) providing those questions to the student for a week to reflect and write about, and 3) the author team joining together for an audio recorded session where the reflections were shared and additional questions arose for more context and insight. After each session, a new set of questions were generated and the cycle would start over again. Informed by studies using autoethnographic methods, we incorporated ``provocations'' during parts of the question/prompt formulation stage to instigate critical reflection on issues that might be overlooked \cite{bates2020integrating, pangrazio2017exploring}.

Five reflective cycles took place over the spring and summer of 2020, resulting in over 5 hours of recorded audio data and transcripts. A model of the reflective cycle process can be seen in Figure 1 of the supplementary materials. The questions posed and generated from the reflective cycles were broad in the first sessions and gradually became more focused and specific (see Figure 2 of the supplementary materials). This methodology is both a potential pedagogical tool as well as qualitative research method. As a method, the reflective cycle is an approach that shifts the research from observational to participatory, and having the student as co-author extends her participation to ownership of her own story as told in this academic paper. 

\subsection{Contextualizing Salinas, CA}

The site of this study, Salinas, is a rural community located in northern California, approximately 70 miles south of Silicon Valley. Nicknamed the ``Salad Bowl of the World'', Salinas is home to a multi-billion dollar agricultural industry. Nearly 79\% of the residents are Latinx (US Census) and make up the vast majority of local agricultural laborers. The agricultural labor community based in and around Salinas has long been a powerful political, cultural, and social force through activism (including protests with César Chávez and the United Farm Workers) and creative expression such as Chicano theater (El Teatro Campesino and Luis Valdez). Like other rural communities in the US, Salinas has few formal opportunities for students to learn CS or AI \cite{center2019computer}, and in general education levels in Salinas are low\footnote{Forbes: Salinas is second least educated city in US (\url{https://bit.ly/3r1kNp2})}, though there are some signs that this trend is reversing\footnote{Monterey Herald: Monterey County schools showing improvement according to state database (\url{https://bit.ly/2T0gP3m})}. Salinas also has many social problems with violence, drugs, and gangs, and in 2016 had the highest youth murder rate in California \cite{langley_sugarmann}. In recent years, Salinas has become a global hub for ag-tech, but there has been limited work in creating inclusive spaces for community-members and laborers to collaborate in technology building and design.

\section{Autoethnographic Case Study Through a Reflective Cycle Process}
The following section presents key excerpts and findings from the autoethnographic reflective cycle process. The events follow the timeline included in Figure 3 of the supplementary materials. The timeline focuses on crucial moments of her learning pathway from middle school through high school and the progression of her interests in environmental science, CS, and AI.

\subsection{Backdrop: Living and Learning in Salinas}
Many residents of Salinas, including the student and her family, are first or second generation immigrants who came to work in the local agricultural industry. The student describes her family’s move from Mexico and the work opportunities of her grandfather in the Bracero Program for agricultural workers that led to that happening:

\begin{quote}
\textit{When my parents immigrated … we moved to Salinas, and I was born. The reason why they moved to Salinas was because of my grandpa. He was part of the Bracero program...a lot of contracted men from Mexico came to work in the strawberry fields in Salinas…\textbf{(Reflective Cycle 1)}}
\end{quote}

Growing up in a family with five siblings, the student notes the language and formal education struggles faced by her older siblings who were born in Mexico. All but two of her siblings had some formal schooling in Mexico, and the student’s entire K-12 education happened in Salinas public schools. Educational and social opportunities for Mexican heritage students can vary a lot by neighborhood in Salinas, an important formative perspective for her. She outlines her understanding of the importance of education and speaking English as it relates to opportunities in the USA:

\begin{quote}
\textit{My first two siblings struggled and they didn't learn English [at K-12 school]. If you don't have education, or you don't have the ability to speak English your opportunities and what you can do are very limited. I remember there was a lot of gang related violence...a lot of policing, a lot of focus was given to that because violence and drugs was very prominent in Salinas, and even within my neighborhood. \textbf{(Reflective Cycle 1)}}
\end{quote}

These combined factors of her family’s origin in Mexico, close family involvement in agricultural work in Salinas, the value placed on education, and growing up in a Mexican-American community helped shape her identity and values. These factors provided important grounding in her learning pathway, as seen later in her applications of AI.

\subsection{First Learning CS} 
The student’s recount of her first STEM opportunities that helped initiate her pathway to CS and AI took shape through an informal after school coding club CoderDojo. CoderDojo is a worldwide community of 2,315 ``free, open, and local'' programming clubs for young people designed to reduce participation gaps in CS \cite{sheridan2016exploration}. She describes her fortuitous initial participation in a nearby chapter of this organization, CoderDojo Hartnell College which included a staff of college students from California State University, Monterey Bay:

	\begin{quote}
\textit{I have a nephew who lives in Salinas too, and my sister wanted to enroll him in a program. She asked me if I wanted to go [to CoderDojo] with him and I said sure so I just tagged along. I thought it would be cool for us to go together. I didn't really know what it was about. The only intention I had was to spend time with my nephew, and then it turns out I'm really interested in this. I was starting to learn about very basic technology stuff like HTML and websites and little robots. So that's how I got into it. \textbf{(Reflective Cycle 2)}}
\end{quote}

The author’s original intention in participating in CoderDojo was not to learn coding specifically but rather to spend time with family. The structure of CoderDojo, like other out of school learning programs, allowed the student to discover her interest in CS in a collaborative environment where she was able to interact with students her age and other ages like her nephew. Informal, out of school programs like CoderDojo created interest based learning opportunities for the author, which in turn may have reinforced her identification and involvement with the discipline. Prior research has found that extracurricular science activities is important in cultivating more positive attitudes towards science, particularly for underrepresented learners, including women in CS \cite{mohr2014developing}.

Spending time at CoderDojo helped develop her early interests, but it was not until she was able to have a computer of her own as a teenager that she started to take her interests more seriously:
\begin{quote}
\textit{I remember I bought a computer, which gave me access to the internet, of course, and so there I was able to learn more about coding, these types of programs and technology. I think having access to a computer was really pivotal in my decision to continue learning and immersing myself in STEM. Without having access to a computer, I would have never gone to do programs or be recognized for my efforts in STEM. I was turning 13 or 14 and thinking about my future. \textbf{(Reflective Cycle 2)}}
	\end{quote}

Access to a computer at home enabled her participation on the internet and in turn awareness of future learning opportunities beyond regular schoolwork. CoderDojo also presented an early opportunity to use her laptop to learn about CS in interesting ways outside of school. In this sense, it was the coupling of access and understanding of how to navigate and identify opportunity that led to continued CS learning.

Despite some informal out-of-school activities and experiences within Salinas, the student outlined how there was often not enough opportunity to develop her growing interests. This desire to learn expanded her pursuit of CS opportunities outside of her local geographic community. Her first such experience, which her AVID\footnote{Advancement Via Individual Determination; Federally-funded education program to help students prepare for college eligibility and success, primarily first-gen students} teacher initially told her about, is described as follows:

	\begin{quote}
\textit{UC Santa Cruz had this exposition for two days in the School of Engineering and we went there and learned about Legos, like, coding Legos and learning more about mini robots. So that's technically the very, very first one. But the reason why I don't really talk about it was because it wasn't very influential. \textbf{(Reflective Cycle 2)}}
	\end{quote}

When hearing this, we wondered what types of experience merited being ‘influential’ in the student’s life, and whether there might be takeaways and reflections related to CS access and education that could be learned from this. The student later encountered and followed female role-model influencers on social media, which helped shape her continued participation. Importantly, these interests led to her involvement with Kode with Klossy, a summer coding program intended for young women, and spurred her interest in learning outside of school.

\begin{quote}
\textit{Kode with Klossy influenced me a lot. It made me realize that I could do more in addition to attending CoderDojo every Saturday and Sunday, that it's actually something that I can do outside of middle school and possibly pursue in the future. At Kode With Klossy, I felt a sense of community for the first time, because in middle school, it seemed like a hobby or something I thought was interesting. I was an incoming freshman at Salinas High School ... and one of the youngest at the program. Kode with Klossy was important to me because there were girls who were going to study CS at Berkeley, and through them and hearing all these girls from different parts of the country coming together and discussing CS gave me a lot of insight into the field and what I could do if I wanted to pursue that … I started to learn the more technical side of computer science and code more seriously. I created a website with other girls [as a project for the program]... \textbf{(Reflective Cycle 2)}}
\end{quote}

The student’s experience at the summer program exposed her to more extracurricular possibilities in CS and inspired her to continue learning CS, even if that meant leaving her hometown. It was her sense of belonging in this community that also made her think about the representation of people like her in STEM fields. Her time with Kode with Klossy emphasized the importance of personal identity in computing and education, in this case through her gender identity. Given this fundamental connection, we wondered how personal identity could also shape ideas of how to use computing technology to define and solve social issues. These experiences led to our next set of questions around the importance of her interpersonal relationships.

\subsection{Interpersonal Relationships}
We next explored her interpersonal relationships and their importance to her learning pathway. The student shared her experiences as part of AI4ALL, an organization that aims to increase diversity and inclusivity in AI. After participating in the AI4ALL summer program, the student participated in the AI4ALL Research Fellowship, a primarily remote project based program. Interpersonal relationships from AI4ALL and, as we learned, her family were pivotal in her learning pathway. 

During her time with the AI4ALL Research Fellowship, the student experienced two contrasting cases of mentorship. Her first mentor ``\textit{didn't offer a lot of support}’’ \textbf{(Reflective Cycle 3)} and was difficult to schedule and communicate with. The student depended on this mentor and therefore her learning experience was somewhat derailed. A data scientist who knew one of the administrators of AI4ALL volunteered to be her mentor. This new mentor shared a similar identity to the student, as they both identify as Latinx. She describes the supportive practices of the new mentor:

	\begin{quote}
\textit{… so when I was connected with my second mentor … she was always very helpful. That's why I really loved my experience with her: because she was always willing to help me … She would send me articles; it was a lot of like article sending and a lot of discussion. We met virtually, she would help me with my code, see what was happening, like checking the systems, making sure they're all running, making sure I didn't do something wrong with the programs ... And then once she gave me examples of how to do stuff, she would say, like, ``hey, here are more articles or examples and here is how you can fix these errors…’’ \textbf{(Reflective Cycle 5)}}
\end{quote}

This excerpt illustrates the importance of good mentorship on the student and her learning journey. The student also expressed how she relied on the support that the mentors provided in order to continue on this path. These practices include being available to communicate through a variety of means, expressing willingness to help, and providing concrete constructive examples. We also explored her relationships with family and community in relation her learning pathway:

\begin{quote}
\textit{I identify as a first-generation Mexican American. My family is an immigrant family from a poor village in Mexico and my parents are both immigrants from Mexico. There was a language barrier: my mom and dad only speak Spanish … My siblings are the first ones in our entire family tree to go to college. So there's that educational barrier that's always existed in my family. \textbf{(Reflective Cycle 1)}}
\end{quote}

Clearly, education was important to the author’s family, thus influencing her motivations and pursuits of further formal opportunities. The importance of family for the student in developing and pursuing her interest in CS was generational, including other members of her own generation. Even though no one in her family was directly involved in CS, they pushed her to rely more on outside programs and search further. The cumulative support of her family offered a hope and source of resilience for her to overcome the specific educational barriers they have faced, which also entailed economic, social, linguistic, and geographic barriers. 

Specifically, family and friends helped her overcome logistical deterrents such as transportation, which the student mentioned on several occasions as being a primary barrier. For example, in this passage she describes the process of going from Salinas to the Bay Area to participate in a two day event with AI4ALL during her academic school year:

\begin{quote}
\textit{I went to [San Francisco] to attend one of the AI4ALL workshops and I had to do so much to get there. My friend had to drive me up to San Jose and then from San Jose, I took the Caltrain, and then took Uber to get to the [tech company headquarters]\footnote{Anonymized for privacy}. I had to get a lot of transportation for me to go there to learn about data science, what [tech company] does, and talk to their engineers. I understood what I have to do to get there, but I still do it because I think it's enriching, and it's going to help move me closer to my goal of pursuing higher education and learning about AI. I stayed overnight at my friend's college dorm near San Francisco. So I stayed in their dorm that day and then in the morning, I took [Bay Area Rapid Transit] to [tech company]. After the program ended, I took Caltrain all the way down to Gilroy and then my brother came to pick me up that day. So that was a two day event. \textbf{(Reflective Cycle 3)}}
\end{quote}

Transportation is usually not considered part of learning, but without the support of her family it could have easily ended her participation in AI4ALL and other outside learning opportunities. Her family and friends provided logistical support through transportation and housing, while her own resilience and resistance was also apparent in these passages, these two factors were important to her learning pathway of CS and eventually AI. Tech companies participate in programs like AI4ALL for various reasons, including the goal of improving their image and inculcating a reputation for doing social good. But the previous passage highlights the need for these companies to address the practical barriers to participation, such as those experienced by the student. 

\subsection{Learning and Applying AI}

Her experiences with AI4ALL started her shift from learning CS to specifically learning AI. The student had early interests in environmental science due to her personal relationship with agriculture in Salinas. Her family works in the agricultural industry there and her understanding of the effects of pollutants from agriculture on workers comes from first hand experiences. This section shows the evolution of her interests that led to her using AI to study environmental problems. Importantly, we found that her justice-oriented perspective developed years before she ever learned about AI.
	
The student’s early interests in environmental science served as a starting point for how she was thinking about applying what she learned, beginning with a middle school science fair project ``\textit{researching how worms could help biodegradable plastic decompose}’’ \textbf{(Reflective Cycle 5)}. This experience extended into a more involved environmental science project in high school with the goal of addressing issues of immediate concern to her and her community:
\begin{quote}
\textit{I wanted to investigate Salinas, specifically agriculture in Salinas. I also wanted to do something with farmworking … I chose that because I thought of the science fair project as a way for me to make a difference in like a very small scale like in terms of finding something interesting in my community, or finding something that is interesting and that we could use for the future or that could help me help others. My science fair research project was seeing how [water fleas] would react to agricultural river water and non-agricultural river water. So the idea behind it was seeing how it affected their heartbeat, because they are very sensitive to water toxicity, and the way you can detect [abnormality] is very simple: you measure their heartbeats. And through their heartbeat, you could learn about the water quality and what and how it's affecting them... \textbf{(Reflective Cycle 5)}}
\end{quote}

Both of these projects provided opportunities for her to explore her developing interests in applying science to issues faced by her community. Up to this point she worked with the tools available to her, none of which were related to AI. Once she learned about AI during her time with AI4ALL and spent time interacting with students and engineers from Silicon Valley, she continued making connections between science, technology, and her community: 

	\begin{quote}
\textit{So then the summer of 2017, well even before that, I think it was midpoint of freshman year I started to look for summer programs on the internet, and landed on the Stanford AI4ALL page. In the summer of 2017 I spent two weeks at Stanford, learning about AI with professors, [tech company] engineers, working with PhD students on autonomous vehicles ... It was really empowering...That's where I learned about AI and the importance of women and diversity in technology. \textbf{(Reflective Cycle 2)}}
\end{quote}

After learning about AI through AI4ALL’s educational program, the student participated in a research fellowship. Her work was supported by the second mentor we described previously. She connected her past middle and high school research experience with water pollution to study similar problems in a larger ecosystem but with tools from AI:
\begin{quote}
\textit{My project analyzed the water quality of the Colorado River. I chose the Colorado River because it is very important to wildlife and humans, and also because there's a lot of data. The Colorado River serves 36 million people and endangered wildlife … because the Colorado River is so big, I just worked on a certain part of the river, the Green River located in Utah. The Colorado River is endangered, and at times runs dry. The biggest threat to running dry is outdated water management… I wanted to learn about environmental issues because if we were to get a model that could predict the water quality, we could address and prevent them before they happened, whether that was improving water management or doing something to help the water quality not get so bad, investigating leakage into the river or whether there's some type of pollution happening. Monitoring the water could help detect patterns, help us learn about quality and potentially predict pollution in the future. \textbf{(Reflective Cycle 5)}}
\end{quote}

Her approach to the problem was motivated by her environmental concerns and how pollution affects people and wildlife. While acquiring and developing skills in AI, she learned about what these methods could afford, how to communicate results, and the importance of domain knowledge when applying AI to specific problems. She also learned about generalizability in applying computation, such as using her results to predict future water pollution: 
\begin{quote}
\textit{We used k-means to cluster groups together based on similar properties, like sulfur, nitrate, fluoride … Our data came from the National Water Information System through USGS, it would provide us sheets of data of the amount of the water property that was in that water … I would clean the noise from the models and learn about why to delete certain information, whether that was helpful or useful. Once we learned about how much fluoride and chloride should be in water, for example, we would determine whether parts of the river have good or bad water quality. It's also a spectrum, so I wouldn't be making a very strong statement like ``oh this is harmful or toxic'' It was just more of learning about it and seeing what side of the spectrum it would fall on. A lot of the challenge was understanding [from data and models] what good or bad water quality is. \textbf{(Reflective Cycle 5)}}
\end{quote}

These experiences all happened outside of schools in Salinas yet they allowed her to gain new perspectives on issues that affect Salinas. This helped her connect with her identity, culture, and values in new ways. These opportunities also served to empower her with the confidence and tools to see herself as someone capable of leading positive social change in her community. The student defined social good as using computation to help alleviate water pollution, but other organizations and people in Salinas that run the multi-billion agricultural industry in Salinas align social good through AI robotics and automation.



\section{Discussion, Conclusion, and Future Directions}
This autoethnographic case study provided rich insights into a learning pathway that culminated in applying AI to a social problem (water pollution) that disproportionately affects the community where the student came from. The student also discussed enabling factors and struggles, and how her community support network and cultural knowledge played a role along the way. Overshadowing her learning pathway were the conflicting aims of how technology and AI were being used in her community that ran contrary to her community improvement oriented ideas of how to use technology. Although in many ways her story is one of success and perseverance, giving more students from underrepresented communities opportunities to participate and thrive in this discipline will require attention to the unique circumstances of students in a given community. This includes considering the ways that schools can function as opportunity brokers for students, the importance of family, and practical considerations such as transportation and lodging at out-of-school programs. Importantly, these connections and insights were also generated by the student in her capacity as a co-author of this paper

\subsection{Applying and Extending the Learning Pathways Framework}
By adopting elements of the learning pathways framework to the student’s experience, we gained significant insight into one person’s individual journey learning CS and AI. Specifically, we learned how she navigated learning opportunities as well as how she accessed resources, connections, and communities throughout the years. We learned how her interest in this discipline was sustained through application of her skills to contexts that were meaningful to her and her community, and how supportive women mentorship increased her self-confidence. Further, we identified key moments where the student expressed agency on her learning journey and sought out resources for sustained engagement: 1) finding Kode with Klossy and its online community/following, 2) seeking out compatible mentorship relations, and 3) continuing to find ways to apply her knowledge and skills towards a project that addressed a community need (water pollution).

Our work methodologically branches from the learning pathways framework by directly engaging and hearing from a student through self-narration, reflection, and personal sensemaking. By centering the student’s voice throughout all stages of the process, this paper gained additional depth through the direct reflection of her experiences. For example, through this process the student learned not just CS but also about herself and the backgrounds, experiences, and disparities of and in contrast to the other girls in the programs. Therefore, the strengths of these programs for the student were almost as much about her personal growth and development of social perspectives as it was to learn the content. Thus, we advocate for more participatory and autoethnographic-based studies looking at the experiences of students, hearing from them firsthand and observing, listening, and partnering with them as researchers in the process. 

\subsection{Nodes and Networks of Support}
Along this journey, family, friends, and mentors played key roles throughout, even though they often did not communicate to each other while they helped her. In fact, most of the time the individual nodes of her support network were completely unaware of who else was helping her at a given time. Her support network was fragmented and disorganized and required navigating multiple one-to-one relationships along the way. As indicated multiple times through the case study, the student’s support network consisted of key enablers both logistically as well as educationally and socially. But the lack of a unified front in her support network shows an important and potentially overlooked aspect of the types of networks and relationships this and other students have: although related research is often situated in a ``community’’, or a ``school’’, or with a ``family’’, the individuals that make up these collective nouns are individually important for students' learning pathways. 

It is clear that on her journey she developed rich relationships developed on an individual basis, each rooted in particular activities, purposes, histories, and within geographies. She took on the role of organizing and synthesizing this complex web of actors while also learning more about her own interests and self as a female Mexican-American youth. This leads to a crucial question for educational stakeholders: How might they be more collaboratively unified around understanding the needs, wants, struggles, and journey of a student? This navigational and support challenge prompts reflection on designs and systems for knowledge sharing centered around sustaining youth interest and learning. These systems would benefit students more if the specific relational knowledge gained from family members, teachers, and mentors were better incorporated. Facilitating stronger inter-connectivity in a student's network could better support them.

\subsection{Towards a Holistic Approach to Equity for Out-of-School Programs}
Identity-based equity strategies were effective for the student. These included Kode with Klossy and AI4ALL which placed an emphasis on encouraging female participation within computing. These findings align with prior work on female participation in CS and STEM, specifically as it relates to same-gender same-race mentors, role models, and peers pursuing STEM \cite{kricorian2020factors}. But, the case study also made it clear that these issues should extend to logistics and practical barriers. In this sense, gaining acceptance into these programs is usually not enough to ensure equitable participation. Transportation and lodging emerged as critical requirements to participate. We saw this in the student’s described difficulties of scheduling transportation to and from programs. Gathering more qualitative feedback from students to understand what barriers emerge when they get accepted could help programs adapt to the specific needs of students.

However, implementing and designing equity oriented strategies could begin before the students ever show up, and equity-oriented strategies must be involved from start to finish and during a program. Based on the student’s experiences described above, a more holistic approach to how programs are positioned (language and value alignment), how they are run (who runs it), their logistics (providing resources to fully participate), how curriculum is modeled (culturally relevant), and how they foster continued participation within CS could enable more equity. Programs could collect feedback from participants to allow them to describe their experiences while participating. Educational programs should carefully consider all of these aspects.

\subsection{For Good and For All, but For Whom?}
``For good’’ and ``for all’’ are terms increasingly used by programs, initiatives and technologies, but who benefits from these ``goods’’? The “for all” is also an explicit invitation to students that might not feel welcome in CS/AI to participate and learn, and in this case they appeared to be successful in delivering an educational experience “for all” from the focal student’s perspective. However, this paper also identified important limits to these programs and their scope. These limits do not suggest limits to the educational efficacy of these types of programs, but parsing the different definitions of “for all” and “social good” could help clarify the roles and relationships of these programs with respect to schools, students, broader communities, and social justice oriented computing. Crucially, and simply put, programs adopting the language of “social good” does not mean that they are using a universally accepted definition of social good; adopting “for all” language does not mean the problems of access and opportunity are solved.

New CS/AI technologies are in development to replace farmworkers, complicating their already unstable political and economic statuses in the US. For agricultural businesses, potential increases in efficiency and scalability of food production are prioritized over potential job displacement. In contrast, the student’s approach to ``social good’’ was shaped by her experiences and those of family and community members. In order to re-imagine better futures, CS/AI programs could invite local organizations to help design projects rooted in community wants and needs. The different actors and organizations involved with the student's learning pathway were important factors in how she she identified a social problem and studied it with AI. Therefore, any application of AI should consider the personnel involved and how their personal identity, including and beyond demographic categories, shapes their perceptions of social good.

\subsection{Implications and Suggestions for CS Education Stakeholders}
We a few key takeaways and provocations for CS education stakeholders:

\begin{itemize}
    \item Developing culturally responsive/relatable/appropriate computing projects and assignments that also reflect and involve the local community
    \item Centering ethics and social-justice oriented practices in all CS education
    \item Positioning the students as central and agentic as they apply AI to problems that are relevant to their local contexts and communities
    \item Building supportive CS learning communities where students feel safe and their identities are well represented by other students
\end{itemize}

\subsection{Limitations and Future Studies}
The student’s journey might have been described differently if written contemporaneously. Given the proximity of Salinas to Silicon Valley, it would be useful to have studies like this in other regions, including more remote areas. To better understand and compare student experiences, multiple studies like this could take place with more students.

\begin{acks}
The authors thank the anonymous reviewers for their helpful feedback. The authors also thank Rubén A. González for feedback that helped sharpen the paper. AJ Alvero is financially supported by the Stanford Diversifying Academia, Recruiting Excellence (DARE) Fellowship. 
\end{acks}

\bibliographystyle{ACM-Reference-Format}
\bibliography{refs}


\end{document}